\documentclass[aps,prl,twocolumn,superscriptaddress,showpacs]{revtex4}

\usepackage{amsmath}
\usepackage{graphicx}

\begin{document}

\title{Deconfinement in a  2D  optical lattice of
coupled 1D boson systems}

\author{A.~F. Ho}
\affiliation{School of Physics and Astronomy,  The University of Birmingham,
Edgbaston, Birmingham B15 2TT, UK.}
\author{M.~A. Cazalilla}
\affiliation{Donostia International Physics Center (DIPC),
Manuel de Lardizabal 4, 20018-Donostia, Spain.}
\author{T. Giamarchi}
\affiliation{University of Geneva, 24 Quai Enerst-Ansermet,
CH-1211 Geneva 4, Switzerland.}

\begin{abstract}
We show that a two-dimensional (2D) array of 1D
interacting boson tubes has a deconfinement transition between a
1D Mott insulator and a 3D superfluid for commensurate fillings
and a dimensional crossover for the incommensurate case. We
determine the phase diagram and excitations of this system and
discuss the consequences for Bose condensates loaded in 2D optical lattices.
\end{abstract}
\maketitle

Loading optical lattices  with  ultracold atoms is providing new
types of many-body systems as well as new ways to look at  hard
problems such as quantum phase transitions in strongly interacting
systems. The recent observation of the superfluid-Mott insulator
transition~\cite{greiner_mott_optical} has demonstrated the large
degree of tunability offered by these setups.  This will allow the
exploration of e.g. spin-1 atoms, fermions, and Bose-Fermi
mixtures, where  strong interactions may lead to novel quantum
many-body states. They offer also the possibility to realize
anisotropic traps and thus to explore the physics of interacting
particles in reduced dimensions
\cite{schreck_mixtures_optical,richard_1dbec_momentum}. Recently,
two dimensional (2D) optical lattices have been realized, where at
each lattice site, a 1D  tube of ultracold atoms is
trapped~\cite{greiner_2dlattice,moritz_bec_1d_coupled}. Such
systems thus  provide  a unique way to obtain strongly interacting
1D Bose gases, of which the Tonks gas is the best known example
(see e.g.~\cite{Dunjko_1Dbec,cazalilla_correlations_1d}).

In 1D, interactions cause more drastic effects than in
higher dimensions and lead, both for fermions and bosons,
to a very peculiar state known as Tomonaga-Luttinger liquid (TLL)
\cite{gogolin_1dbook,giamarchi_1dbook,cazalilla_correlations_1d}, which by now
is theoretically well understood. However, despite the intense
theoretical and experimental activity in fermionic systems or spin
chains \cite{giamarchi_1dbook}, much less is known when a collection of such
1D systems is coupled. In general, one can expect a dimensional
crossover where the system goes from a TLL at high temperature to an
anisotropic 3D system with more conventional properties (e.g. a
Fermi liquid for fermions, a superfluid for bosons) at low temperature.

The transition is even more drastic and less well understood when,
due to the existence of  an axial periodic potential (thereafter
called ``Mott potential''), the 1D system becomes a Mott insulator
(MI)~\cite{haldane_bosons,giamarchi_mott_shortrev,buchler_cic_bec}.
Quite generally, the competition between Mott localization and
superfluidity leads to interesting
phenomena~\cite{fisher_boson_loc}. For coupled 1D chains the
competition between the Mott insulating physics and  interchain
tunneling changes the dimensional crossover into a deconfinement
transition where the system goes from a 1D insulator towards a 3D
metal or superfluid. For fermions this transition remains a
theoretical challenge \cite{biermann_oned_crossover_review} and
has been observed in coupled chain systems such as the organic
conductors \cite{vescoli_confinement_science}. For bosons, much
less is known even though the problem has been tackled in the
context of spin chains ({\it i.e.} hard core bosons)
\cite{schulz_coupled_spinchains} or for a bosonic ladder
\cite{donohue_deconfinement_bosons}.  In general, however,
comparison between theory and experiments is rendered difficult by
the complex nature of the solid-state materials.

\begin{figure}[t]
 \centerline{\includegraphics[width=8cm]{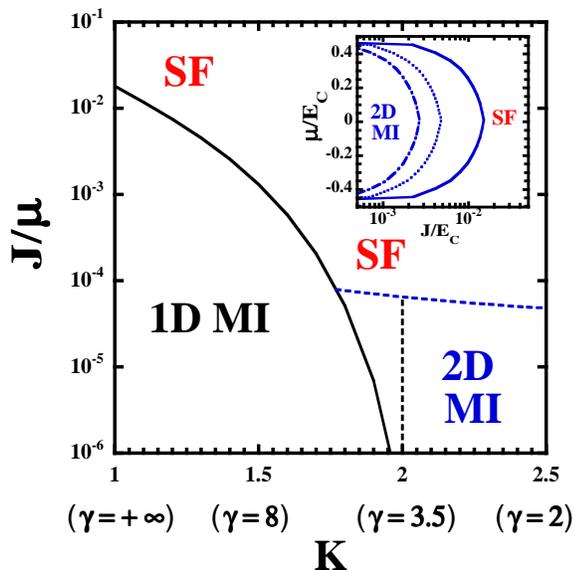}}
 \caption{\label{fig1} Zero-temperature phase diagram of a 2D lattice of
coupled 1D  boson systems. The values on the vertical axis are
defined up to a factor of order unity. $K$ is the TLL parameter.
The corresponding values of $\gamma = Mg/\hbar^2 \rho_0$ are also
given. A periodic potential of amplitude $u_0$, commensurate with
the density $\rho_0$, is applied longitudinally ($u_0\approx 0.03
\: \mu$ for $K=1$  and $u_0 \approx 0.02\: \mu$ for $K=2.5$).  For
infinite 1D systems only two phases exist: a 1D Mott insulator (1D MI)
and a 3D (but anisotropic) superfluid (SF). For finite tubes,
a 2D Mott insulator (2D MI) can exist for $J/\mu$
below the  horizontal dashed curve (we have taken  $N_0
= 100$ atoms per 1D tube). The vertical  dashed line (schematic)
indicates that the transition to the 1D MI should take place for
$K\simeq 2$.  The insert is the phase diagram of a 2D optical
lattice of finite 1D systems for $u_0 = 0$  and $N_0 = 100$, as a function of
chemical potential $\mu$ and the hopping $J$, and for various values of $K$:
$K = 1$ (continuous curve), $K = 2$ (dotted curve), $K= 4$ ($\gamma = 0.71$, dashed-dotted
curve). }
\end{figure}

In this paper, motivated by the \emph{clean} and \emph{tunable}
realization of  coupled 1D gases offered by 2D optical
lattices~\cite{greiner_2dlattice,moritz_bec_1d_coupled}, we
investigate their phases and  phase diagram (see Fig.~\ref{fig1}).
For the lattice without the axial Mott potential, we
show that the system exhibits, as a function of temperature, a
dimensional crossover from a strongly anisotropic normal gas to a
3D  Bose-Einstein condensate (BEC) or superfluid, which can have a
large quantum depletion. For this 3D superfluid (SF), we obtain
the condensation temperature, the zero-temperature condensate
fraction, the excitation spectra, and the momentum distribution in
the thermodynamic limit. When an axial Mott potential is present, we
show that its tendency to lead to a Mott insulator competes with
the Josephson coupling between the 1D tubes, which tries to
delocalize the atoms. Changing the hopping amplitude between the tubes
drives the deconfinement transition, and we determine its
properties and the phase diagram. In addition, since in the
experiments~\cite{greiner_2dlattice,moritz_bec_1d_coupled} the 2D
optical lattices are confined  in a harmonic trap,
we have considered the effect of finite-size tubes. For small
enough hopping a 2D lattice behaves as a 2D array of Josephson
junctions which can undergo a transition to a 2D Mott insulating
state where tunnelling between tubes is suppressed by the
``charging energy'' of each tube.

We consider a 2D square lattice of 1D tubes of length $L$
containing $N_0 = \rho_0 L$ bosons with  Hamiltonian:
\begin{eqnarray}
 H &=& \sum_{\bf R}   \int^L_0 dx\: \left[ \frac{\hbar^2}{2M}
 \partial_x\Psi^{\dag}_{\bf R}(x) \partial_x\Psi_{\bf R}(x)
 + u(x)\rho_{\bf R}(x) \right] \nonumber \\
 &+&  \sum_{\bf R}  \frac{1}{2} \int^{L}_{0} dx \int^{L}_{0} dx' \,
 v(x-x')
 \rho_{\bf R}(x) \rho_{\bf R}(x') \Big] \nonumber \\
 &-& \frac{J}{2} \sum_{\langle{\bf R},{\bf R}' \rangle} \int^L_0 dx
 \left[ \Psi^{\dag}_{\bf R}(x) \Psi_{\bf R'}(x) +{\rm h.c.}
 \right].\label{ham1}
\end{eqnarray}
$\Psi_{\bf R}(x)$ is the bosonic  field operator at $x$ (the axial
coordinate), and at the lattice site ${\bf R}=(m,n) b$, where
$m,n$ are integers and $b$ the lattice parameter; $\rho_{\bf R}(x) =
\Psi^{\dag}_{\bf R}(x) \Psi_{\bf R}(x)$ the density operator and
$u(x)$ is the Mott potential.  
For cold bosonic atoms confined to
1D,  the interaction  $v(x) = g_{\rm 1D}
\: \delta(x)$, where $g_{\rm 1D} =  2\hbar^2
a_{3D}[(1- C a_{3D}/\sqrt{2}\ell_{\perp})M \ell^{2}_{\perp}]^{-1}$
\cite{olshanii_potential_bec},  being $a_{3D}$  the 3D scattering length,
$C \simeq 1.4603$,  and $\ell_{\perp} \simeq b V_{0\perp}^{-1/4}/\pi$ the
transverse oscillator length~\cite{greiner_mott_optical}
 ($V_{0\perp}$ is the strength of the
 transverse optical potential  in units 
of the recoil energy $E_R = \hbar^2 \pi^2/2M b^2$).
In the last term of (\ref{ham1}), 
which describes the Josephson coupling of the 1D tubes,
$\langle{\bf R},{\bf R'}\rangle$ stands for  sum over nearest neighbors.
Furthermore, we assume  that the 2D optical
lattice  is deep enough for the hopping
 $J \ll \mu$,  $\mu$ being the 1D chemical potential.
We first focus on the thermodynamic properties ({\it i.e.}
$N_0\to \infty$, and $L\to \infty$); finite-size and trap effects will be
considered towards the end.

For the low-temperature properties of (\ref{ham1}), it is
convenient to use the so-called bosonization technique
\cite{gogolin_1dbook,giamarchi_1dbook,cazalilla_correlations_1d}.
Using $\Psi_{\bf R}(x) \simeq \sqrt{A \rho_0}\: e^{i\phi_{\bf
R}(x)}$ and $\rho_{\bf R}(x) = \left[\rho_0+\partial_x\theta_{\bf
R}(x)/\pi\right] \sum_{m} e^{2im[\theta_{\bf R}(x)+\pi \rho_0
x]}$, Eq.~(\ref{ham1}) becomes
\begin{eqnarray}
H_{\rm eff}&=&\frac{\hbar v_s}{2\pi} \sum_{\bf R} \int^{L}_{0}
dx\: \left[\frac1K \left(\partial_x\theta_{\bf R}(x)\right)^2
+ K  \left(\partial_x\phi_{\bf R}(x)\right)^2  \right]\nonumber\\
&+& \frac{\hbar v_s g_u}{2\pi a^2} \sum_{\bf R}
 \int^{L}_{0} dx \: \cos \left(2\theta_{\bf R}(x) + \delta \pi x \right)\nonumber\\
&-& \frac{\hbar v_s g_J}{2\pi a^2} \sum_{\langle {\bf R}, {\bf R}' \rangle} \int^{L}_{0} dx\:
\cos\left(\phi_{\bf R}(x) - \phi_{\bf R'}(x)  \right) \label{ham2}
\end{eqnarray}
where $\delta = 2\rho_0 - G/\pi$ is the mismatch between the
density $\rho_0$ and the periodicity of  $u(x) = u_0 \cos(G x)$.
The dimensionless couplings $g_J = 2\pi J (\rho_0 a)^2 A/\hbar v_s
\rho_0$, and $g_u = 2\pi u_0 (\rho_0 a)^2/\hbar v_s \rho_0$, where
$a  \approx \hbar v_s/\mu$ is the short-distance cut-off and $A
\approx (K/\pi)^{1/2K}$~\cite{cazalilla_correlations_1d}. Results
and asymptotic expressions for  $K$ and the sound velocity $v_s$
in terms of $\gamma = M g_{1D}/\hbar^2 \rho_0$ can be found in
Ref.~\cite{cazalilla_correlations_1d}. Interestingly, an
anisotropic version of the Bose-Hubbard
model~\cite{fisher_boson_loc} (\emph{i.e.} with axial hopping $J_x
\gg J$) also leads to Eq.~(\ref{ham2}) ~\cite{ho_dimcross_long}, but
the relationship of the microscopic parameters (namely
$J_x, J$ and $U$) to $K$, $v_s$, $g_u$ and $g_J$ is not easy to
estimate. But if instead $K$ and $g_u$ are regarded as 
phenomenological parameters, our results directly apply to this
model as well.

First we consider the case $u_0 = 0$ ({\it i.e.} $g_u = 0$). This
includes the incommensurate case $\delta \ne 0$ for which the
periodic potential $u(x)$ has no effect at low
temperatures~\cite{haldane_bosons,giamarchi_mott_shortrev,
buchler_cic_bec}. In  the absence of tunnelling, the system is
then a set of  isolated TLL's with no true condensate and a
power-law decay of phase correlations~\cite{haldane_bosons}.
Tunneling   thus induces,  at a temperature $T_c$, a dimensional
crossover from the TLL behavior at $T>T_c$, towards a 3D
superfluid or BEC at low temperatures. To study the crossover, we
treat the tunneling term in a mean-field (MF) approximation,
taking $\psi_0 = \langle \Psi_{\bf R}(x) \rangle$ to be the
(square root of the) condensate fraction. The problem then reduces
to a sine-Gordon (SG) model~\cite{schulz_coupled_spinchains,
ho_dimcross_long}, describing an effective 1D system. Above
$T_c$, $\psi_0 = 0$.   Therefore, $T_c$ is found by the existence
of a non-zero solution for $\psi_0$. We get, in agreement with
\cite{efetov_coupled_bosons},
\begin{eqnarray} \label{TcMFT}
 \left(\frac{2 \pi T_c}{\hbar v_s \rho_0}\right)^{2-1/2K}
 = f(K) \frac{4 J}{\hbar v_s \rho_0} ,
\end{eqnarray}
where $f(K)= A \sin\left(\frac{\pi}{4K}\right)
B^2\left(\frac{1}{8K}, 1- \frac{1}{4K}\right)$ and $B(x)$ is the
Beta function.  Below $T_c$, $\psi_0$ is found by minimizing the
free energy energy density. At $T=0$ one minimizes ${\cal E}_{\rm
MF}(\psi_0) = 4 J \psi^2_0 -\frac{\Delta_s^2(\psi_0)}{4 \hbar v_s}
\tan{\frac{\pi}{2(8K-1)}}$, where the second term is the SG model
energy density \cite{zamolodchikov_energy_sg}, and
$\Delta_{s}(\psi_0)$ the soliton gap (the excitations of the SG
model  are gapped solitons and breathers
\cite{gogolin_1dbook,giamarchi_1dbook,thacker_rmp}). Thus we find
that the condensate fraction $\psi^2_0(T=0) \sim \rho_0
(J/\mu)^{1/(4K-1)}$ whereas the soliton gap $\Delta_s \sim  \mu
(J/\mu)^{2K/(4K-1)}$ \footnote{The precise relationship
\cite{zamolodchikov_energy_sg} between $\Delta_s$ and the coupling
of the SG term  allows us to obtain the full dependence of
$\psi_0$ on $K$ and the non-universal ratio $A/(\rho_0
a)^{1-1/2K}$. The full expression will be given elsewhere
\cite{ho_dimcross_long}.}.

In the broken-symmetry (superfluid) phase the system
possesses two modes. The Goldstone mode ({\it i.e.} oscillations of
the order parameter phase) is gapless as a consequence of (global)
gauge invariance. The other mode corresponds to oscillations of the order parameter
amplitude. To obtain these modes we compute the gaussian
 fluctuations around the MF solution
(random-phase approximation, RPA). Using known properties of
correlation functions of the  SG model
\cite{gogolin_1dbook,giamarchi_1dbook,zamolodchikov_energy_sg},
and within the single-mode approximation (SMA: taking into account
only the lowest breather mode) \cite{schulz_coupled_spinchains},
we obtain:
\begin{eqnarray} \label{RPAgoldstone}
 \omega^2_{-}(q,{\bf Q}) &=&  v^2_s q^2 + \frac{ \Delta^2_1}{2\hbar^2}\: F({\bf Q}) \\
 \omega^2_{+}(q,{\bf Q}) &=&  \Delta^{2}_{+}  + v^2_s q^2 + \left(\frac{\Delta^2_{+}/2\hbar^2}{8K-2}\right)
 F({\bf Q}) \label{RPAgapped}
\end{eqnarray}
where $\Delta^2_{+} = \Delta^2_2 (8K-2)/(8K-1)$ and $q$ (resp.
${\bf Q}$) is the momentum along (resp. perpendicular to) the chains.
$\Delta_n = \Delta_s \sin \left(n\pi/2(8K-1)\right)$ is the energy
gap for the $n$-th  breather  ($n=1,2,\ldots$), and $\displaystyle
F({\bf Q}) =  \sum_{j=y,z}\left(1- \cos Q_j b\right)$.

For weakly interacting bosons ($K\gg 1$), the use of SMA may be
questioned because  the breather excitations of the mean-field SG
model
proliferate~\cite{gogolin_1dbook,giamarchi_1dbook,thacker_rmp}.
Therefore, we have employed a variational approach previously used
for coupled spin chains \cite{donohue_thesis}, and found the same
dependence on $J/\mu$ for the dispersion as in
(\ref{RPAgoldstone}). With this approach, we have also calculated
the  momentum distribution  of the superfluid phase (in the
absence of a trap). At $T=0$ and small momenta ($\ll \pi/b$),
$\Psi_{\bf R}(x) \simeq \rho^{1/2}_0 \, e^{i\phi_{\bf R}(x)}$ and
$\omega^2_{-}(q,{\bf Q}) \simeq  v^2_s q^2 + v^2_{\perp} {\bf
Q}^2$,  ($v_{\perp} \sim
 \mu b (J/\mu)^{2K/(4K-1)}/\hbar$, cf. Eq.~(\ref{RPAgoldstone})),
 we find   an anisotropic generalization of the Bogoliubov result~\cite{pitaevskii_becbook}:
\begin{eqnarray}
 \frac{n(q,{\bf Q})}{|w({\bf Q})|^2} \simeq  \psi^2_0 \delta({\bf Q}) \delta(q) +
 \frac{\pi b^2 \psi^2_0/2 K} {\left[q^2 + \left(v_{\perp}{\bf Q}/v_{s}\right)^2
 \right]^{1/2}},
\end{eqnarray}
which is valid for arbitrary interactions within the tubes. $w({\bf Q})$
is the Fourier transform of the Wannier orbital, which varies slowly
with $\bf Q$.

We next consider  the case of  a finite axial Mott potential, {\it
i.e.} $u_0\neq 0$ and $\delta \simeq 0$. In the absence of
tunneling, the individual tubes would be 1D MI's: even for a weak
Mott potential, an integer number of bosons are localized to each
well of the Mott potential
\cite{haldane_bosons,giamarchi_mott_shortrev,buchler_cic_bec}.
Increasing the tunneling will thus lead to  deconfinement ({\it
i.e.} 1D MI to 3D SF transition). To study the competition between
the the Mott potential ($g_u$) and the Josephson coupling ($g_J$),
we have used  the renormalization group (RG) method. Since the
Hamiltonian in~(\ref{ham2}) defines an effective field theory, its
couplings ($K$, $g_u$, $g_J$, etc.) depend on the cut-off energy
scale, which is typically set by the temperature for $T < \mu$.
Thus, as the temperature is decreased, the couplings $K, g_u, g_J$
are renormalized, due to virtual transitions to high energy
states. The nature of the ground state is thus determined by the
dominant coupling as $T\to 0$. Using perturbative RG, we find
the following RG flow equations:
\begin{eqnarray}
\frac{dg_F}{d\ell}  &=& \frac{g^2_J}{K}, \qquad
\frac{dg_J}{d\ell}   = \left(2-\frac{1}{2K}\right)g_J + \frac{g_Jg_F}{2K}, \label{rg2}\\
\frac{dg_u}{d\ell}  &=& \left( 2 - K \right) g_u, \quad \quad
\frac{dK}{d\ell} = 4 g^2_J - g^2_u K^2 \label{rg4},
\end{eqnarray}
where $\ell \approx \ln \mu/T$. The coupling $g_F$ is generated by
the RG  and describes an interaction between bosons in neighboring
tubes; $g_u$ and $g_J$ compete for $1/4 < K < 2$: the relative
magnitude of their initial ({\it i.e.} bare) values determines
which coupling grows faster. The faster growth of one coupling
inhibits the other's growth via the renormalization of $K$:  If
$g_u$ grows faster, $K$ flows towards $0$, leading to the 1D MI,
while if $g_J$ grows faster, $K$ also grows  and leads to the 3D
SF. To estimate the phase boundary between the 1D MI and the 3D
SF,  we integrate these equations (at fixed $g_u(0)$) with various
$g_J(0)$  till the couplings are of order one. The resulting phase
boundary is  the continuous curve in Fig.~\ref{fig1}. Via a
separate mean field calculation at a specific $K$ value (see
\cite{ho_dimcross_long}), we showed that the condensate fraction
$\psi^2_0$ grows continuously from zero at the phase boundary.

Finally we turn to finite-size effects. First the finite extent of
trapped cloud (either longitudinally or transversally) limits the
minimum momenta of the modes. Thus  the energy 
of the lowest modes in the 3D SF phase can be directly estimated 
from our results by using the minimum  available momentum in (\ref{RPAgoldstone},\ref{RPAgapped}). For
instance, for a finite 2D lattice containing 
$M_y \times M_z$ tubes
(\emph{i.e.} an atom cloud of size $L \times M_y b \times M_z b$)
the lowest available momentum is $\sim \pi/(M_{i} b)$. Putting
this value into (\ref{RPAgoldstone},\ref{RPAgapped}) shows that the
frequency of the lowest transverse modes decreases with decreasing $J$. 
This is in qualitative agreement with the 
hydrodynamic analysis of Ref.\cite{pedri_tonks_harmonic}. 
However, unlike this hydrodynamic analysis, our results  apply to the 
strongly interacting regime of the superfluid near the transition 
where depletion of the condensate can
be large (as shown e.g. by the power-law behavior of the
condensate fraction, $\psi^2_0 \sim \rho_0 (J/\mu)^{1/(4K-1)}$).
In addition,  we are also able to   describe the transition
to the Mott insulating regime. Note that  for finite size 
systems the phase transitions described 
above become crossovers. Furthermore, for harmonically 
trapped systems in the MI regimes an 
inhomogeneous state arises where the the SF and MI
coexist~\cite{greiner_mott_optical,buchler_cic_bec}, and $\psi_0$
decreases slowly across the phase boundaries.

A second type of finite size effects comes from the finite size
and hence discrete spectrum of each tube. In the  experiments in
2D optical lattices loaded with ultracold
atoms~\cite{greiner_2dlattice,moritz_bec_1d_coupled}, there are
typically $N_0 \sim 10^2-10^3$ atoms per  tube. At low
temperatures, if $J$ is made very small ({\it i.e.} the lattice
potential very deep), each 1D tube will behave as an atomic
``quantum dot'' characterized by a charging energy~\footnote{This
is analogous to the Coulomb blockade phenomenon in mesoscopic
physics.} $E_C = \hbar \pi v_s/K
L$~\cite{cazalilla_correlations_1d}. This is to be balanced
against the hopping energy $E_J$, which  is not just $J N_0$ as the
hopping amplitude is renormalized by the creation of virtual sound
waves within the 1D tubes. Thus, to study this effect for $u_0 =0$, we
integrate out the phonons perturbatively in $J$ and obtain a quantum-phase model:
\begin{eqnarray}
H_{\rm QP} &=& - E_J \sum_{\langle{\bf R}, {\bf R'}\rangle}
\cos \left(\phi_{0{\bf R}} - \phi_{0{\bf R'}} \right) \nonumber \\
& & +\frac{E_C}{2} \sum_{\bf R} (N_{\bf R}-N_0)^2 -\mu \sum_{\bf R}
N_{\bf R},\label{qp}
\end{eqnarray}
where $N_{\bf R}$ (resp. $\phi_{0{\bf R}}$) is the particle-number
(reps. phase) operator of tube $\bf R$ (hence $\left[N_{\bf
R},\phi_{0{\bf R'}} \right]= i\delta_{\bf R, \bf R'}$). The
renormalized hopping $E_J = E_J(N_0) \simeq J
N^{1-\frac{1}{2K}}_0$. The model~(\ref{qp}) has been extensively
studied in the literature in connection with 2D arrays of
Josephson junctions. It exhibits a 2D MI-SF  at a critical value
of $E_J/E_C$. For commensurate filling $N_0$ ($\mu = 0$), a Monte
Carlo (MC) calculation~\cite{vanotterlo_2djjarrays} gives
$(E_J/E_C)_c \simeq 0.15$. In our case this reduces to $(J/\mu)_c
\simeq 0.3 N^{-3/2}_0$ in the Tonks limit ($K\simeq 1$) and to
$(J/\mu)_c \simeq 0.15 N^{-2}_0$  for weakly interacting bosons
($K \gg 1$). Away from commensurate filling ($\mu \neq 0$), the
critical $J/\mu$ is reduced as shown in the insert of Fig.~\ref{fig1},
obtained using the MC data from Ref.~\cite{vanotterlo_2djjarrays}.
Note that in the 2D MI, only the phase coherence {\it between}
different  tubes  of the 2D lattice is lost. There is no gap
(apart from the finite size gap) to excitations {\it within} each
1D system. This is different from the 1D MI described above, where
all excitations are gapped and any order (even quasi-long range)
in the phase is absent. For finite $u_0$, when $K \gtrsim 2$,
$g_u$ flows to zero and the consideration above applies (hence the
horizontal and vertical dashed lines in Fig.~\ref{fig1}).
Otherwise ($K \lesssim 2$), $g_J$ flows to zero and the 1D MI is
as described above.

The above predictions can be directly tested in 2D optical
lattices~\cite{greiner_2dlattice,moritz_bec_1d_coupled}. In presence of the
Mott potential, the deconfinement transition is of course the
easiest to check. Even without the Mott potential, $\psi^2_0$  and
$T_c$ as functions of $J/\mu$,  and the change in the momentum distribution
from the 1D TLL to the 3D superfluid should be accessible, by releasing the trap
and the optical lattice, then measuring the expansion images.
Finally, the excitation spectra (\ref{RPAgoldstone},\ref{RPAgapped})
of the superfluid  can be measured with Bragg spectroscopy at low
momentum transfer.

We thank  T. Esslinger, A. Nersesyan,
and M. Gunn for useful conversations/correspondence.
AFH and MAC thank the hospitality of the University of
Geneva where this work was started. AFH is funded by EPSRC (UK),
MAC by Gipuzkoako Foru Aldundia (Basque Country). This work was supported in part
by the Swiss National Science Foundation through MaNEP.

{\it Note Added.-} After this Letter was submitted, we received a preprint from
Esslinger \cite{Stoferle04} reporting on experimental results for the system we have
considered here. The reported results are at least qualitatively consistent with our predictions. 

%\bibliography{totphys,bose1d}

\begin{thebibliography}{26}
\expandafter\ifx\csname natexlab\endcsname\relax\def\natexlab#1{#1}\fi
\expandafter\ifx\csname bibnamefont\endcsname\relax
  \def\bibnamefont#1{#1}\fi
\expandafter\ifx\csname bibfnamefont\endcsname\relax
  \def\bibfnamefont#1{#1}\fi
\expandafter\ifx\csname citenamefont\endcsname\relax
  \def\citenamefont#1{#1}\fi
\expandafter\ifx\csname url\endcsname\relax
  \def\url#1{\texttt{#1}}\fi
\expandafter\ifx\csname urlprefix\endcsname\relax\def\urlprefix{URL }\fi
\providecommand{\bibinfo}[2]{#2}
\providecommand{\eprint}[2][]{\url{#2}}

\bibitem[{\citenamefont{Greiner et~al.}(2002)\citenamefont{Greiner, Mandel,
  Esslinger, Hansch, and Bloch}}]{greiner_mott_optical}
\bibinfo{author}{\bibfnamefont{M.}~\bibnamefont{Greiner}},
  \bibinfo{author}{\bibfnamefont{O.}~\bibnamefont{Mandel}},
  \bibinfo{author}{\bibfnamefont{T.}~\bibnamefont{Esslinger}},
  \bibinfo{author}{\bibfnamefont{T.~W.} \bibnamefont{Hansch}},
  \bibnamefont{and} \bibinfo{author}{\bibfnamefont{I.}~\bibnamefont{Bloch}},
  \bibinfo{journal}{Nature} \textbf{\bibinfo{volume}{415}}, \bibinfo{pages}{39}
  (\bibinfo{year}{2002}).

\bibitem[{\citenamefont{Schreck et~al.}(2001)\citenamefont{Schreck, Khaykovich,
  Corwin, Ferrari, Bourdel, Cubizolles, and
  Salomon}}]{schreck_mixtures_optical}
\bibinfo{author}{\bibfnamefont{F.}~\bibnamefont{Schreck}},
  \bibinfo{author}{\bibfnamefont{L.}~\bibnamefont{Khaykovich}},
  \bibinfo{author}{\bibfnamefont{K.~L.} \bibnamefont{Corwin}},
  \bibinfo{author}{\bibfnamefont{G.}~\bibnamefont{Ferrari}},
  \bibinfo{author}{\bibfnamefont{T.}~\bibnamefont{Bourdel}},
  \bibinfo{author}{\bibfnamefont{J.}~\bibnamefont{Cubizolles}},
  \bibnamefont{and} \bibinfo{author}{\bibfnamefont{C.}~\bibnamefont{Salomon}},
  \bibinfo{journal}{Phys. Rev. Lett.} \textbf{\bibinfo{volume}{87}},
  \bibinfo{pages}{080403} (\bibinfo{year}{2001}).

\bibitem[{\citenamefont{Richard et~al.}(2003)\citenamefont{Richard, Gerbier,
  Thywissen, Hugbart, Bouyer, and Aspect}}]{richard_1dbec_momentum}
\bibinfo{author}{\bibfnamefont{S.}~\bibnamefont{Richard}},
  \bibinfo{author}{\bibfnamefont{F.}~\bibnamefont{Gerbier}},
  \bibinfo{author}{\bibfnamefont{J.~H.} \bibnamefont{Thywissen}},
  \bibinfo{author}{\bibfnamefont{M.}~\bibnamefont{Hugbart}},
  \bibinfo{author}{\bibfnamefont{P.}~\bibnamefont{Bouyer}}, \bibnamefont{and}
  \bibinfo{author}{\bibfnamefont{A.}~\bibnamefont{Aspect}},
  \bibinfo{journal}{Phys. Rev. Lett.} \textbf{\bibinfo{volume}{91}},
  \bibinfo{pages}{010405} (\bibinfo{year}{2003}).

\bibitem[{\citenamefont{{Greiner {\it et al.}}}(2001)}]{greiner_2dlattice}
\bibinfo{author}{\bibfnamefont{M.}~\bibnamefont{{Greiner {\it et al.}}}},
  \bibinfo{journal}{Phys. Rev. Lett.} \textbf{\bibinfo{volume}{87}},
  \bibinfo{pages}{160405} (\bibinfo{year}{2001}).

\bibitem[{\citenamefont{Moritz et~al.}(2003)\citenamefont{Moritz, St{\"o}ferle,
  K{\"o}hl, and Esslinger}}]{moritz_bec_1d_coupled}
\bibinfo{author}{\bibfnamefont{H.}~\bibnamefont{Moritz}},
  \bibinfo{author}{\bibfnamefont{T.}~\bibnamefont{St{\"o}ferle}},
  \bibinfo{author}{\bibfnamefont{M.}~\bibnamefont{K{\"o}hl}}, \bibnamefont{and}
  \bibinfo{author}{\bibfnamefont{T.}~\bibnamefont{Esslinger}}
  (\bibinfo{year}{2003}), \eprint{cond-mat/0307607}.

\bibitem[{\citenamefont{Dunjko et~al.}(2001)\citenamefont{Dunjko, Lorent, and
  Olshanii}}]{Dunjko_1Dbec}
\bibinfo{author}{\bibfnamefont{V.}~\bibnamefont{Dunjko}},
  \bibinfo{author}{\bibfnamefont{V.}~\bibnamefont{Lorent}}, \bibnamefont{and}
  \bibinfo{author}{\bibfnamefont{M.}~\bibnamefont{Olshanii}},
  \bibinfo{journal}{Phys. Rev. Lett.} \textbf{\bibinfo{volume}{86}},
  \bibinfo{pages}{5413} (\bibinfo{year}{2001}), \eprint{and references
  therein}.

\bibitem[{\citenamefont{Cazalilla}(2003)}]{cazalilla_correlations_1d}
\bibinfo{author}{\bibfnamefont{M.~A.} \bibnamefont{Cazalilla}}
  (\bibinfo{year}{2003}), \eprint{J. Phys. B, in press (cond-mat/0307033)}.

\bibitem[{\citenamefont{Gogolin et~al.}(1999)\citenamefont{Gogolin, Nersesyan,
  and Tsvelik}}]{gogolin_1dbook}
\bibinfo{author}{\bibfnamefont{A.~O.} \bibnamefont{Gogolin}},
  \bibinfo{author}{\bibfnamefont{A.~A.} \bibnamefont{Nersesyan}},
  \bibnamefont{and} \bibinfo{author}{\bibfnamefont{A.~M.}
  \bibnamefont{Tsvelik}}, \emph{\bibinfo{title}{Bosonization and Strongly
  Correlated Systems}} (\bibinfo{publisher}{Cambridge University Press},
  \bibinfo{address}{Cambridge}, \bibinfo{year}{1999}).

\bibitem[{\citenamefont{Giamarchi}(2004)}]{giamarchi_1dbook}
\bibinfo{author}{\bibfnamefont{T.}~\bibnamefont{Giamarchi}},
  \emph{\bibinfo{title}{Quantum Physics in One Dimension}}
  (\bibinfo{publisher}{Clarendon Press}, \bibinfo{address}{Oxford},
  \bibinfo{year}{2004}).

\bibitem[{\citenamefont{Haldane}(1981)}]{haldane_bosons}
\bibinfo{author}{\bibfnamefont{F.~D.~M.} \bibnamefont{Haldane}},
  \bibinfo{journal}{Phys. Rev. Lett.} \textbf{\bibinfo{volume}{47}},
  \bibinfo{pages}{1840} (\bibinfo{year}{1981}).

\bibitem[{\citenamefont{Giamarchi}(1997)}]{giamarchi_mott_shortrev}
\bibinfo{author}{\bibfnamefont{T.}~\bibnamefont{Giamarchi}},
  \bibinfo{journal}{Physica B} \textbf{\bibinfo{volume}{230-232}},
  \bibinfo{pages}{975} (\bibinfo{year}{1997}).

\bibitem[{\citenamefont{B{\"u}chler et~al.}(2003)\citenamefont{B{\"u}chler,
  Blatter, and Zwerger}}]{buchler_cic_bec}
\bibinfo{author}{\bibfnamefont{H.~P.} \bibnamefont{B{\"u}chler}},
  \bibinfo{author}{\bibfnamefont{G.}~\bibnamefont{Blatter}}, \bibnamefont{and}
  \bibinfo{author}{\bibfnamefont{W.}~\bibnamefont{Zwerger}},
  \bibinfo{journal}{Phys. Rev. Lett.} \textbf{\bibinfo{volume}{90}},
  \bibinfo{pages}{130401} (\bibinfo{year}{2003}).

\bibitem[{\citenamefont{Fisher et~al.}(1989)\citenamefont{Fisher, Weichman,
  Grinstein, and Fisher}}]{fisher_boson_loc}
\bibinfo{author}{\bibfnamefont{M.~P.~A.} \bibnamefont{Fisher}},
  \bibinfo{author}{\bibfnamefont{P.~B.} \bibnamefont{Weichman}},
  \bibinfo{author}{\bibfnamefont{G.}~\bibnamefont{Grinstein}},
  \bibnamefont{and} \bibinfo{author}{\bibfnamefont{D.~S.}
  \bibnamefont{Fisher}}, \bibinfo{journal}{Phys. Rev. B}
  \textbf{\bibinfo{volume}{40}}, \bibinfo{pages}{546} (\bibinfo{year}{1989}).

\bibitem[{\citenamefont{Biermann et~al.}(2002)\citenamefont{Biermann, Georges,
  Giamarchi, and Lichtenstein}}]{biermann_oned_crossover_review}
\bibinfo{author}{\bibfnamefont{S.}~\bibnamefont{Biermann}},
  \bibinfo{author}{\bibfnamefont{A.}~\bibnamefont{Georges}},
  \bibinfo{author}{\bibfnamefont{T.}~\bibnamefont{Giamarchi}},
  \bibnamefont{and}
  \bibinfo{author}{\bibfnamefont{A.}~\bibnamefont{Lichtenstein}}, in
  \emph{\bibinfo{booktitle}{Strongly Correlated Fermions and Bosons in Low
  Dimensional Disordered Systems}}, edited by
  \bibinfo{editor}{\bibfnamefont{I.~V.} \bibnamefont{{Lerner {\it et al.}}}}
  (\bibinfo{publisher}{Kluwer Academic Publishers},
  \bibinfo{address}{Dordrecht}, \bibinfo{year}{2002}), p.~\bibinfo{pages}{81},
  \bibinfo{note}{cond-mat/0201542}.

\bibitem[{\citenamefont{Vescoli et~al.}(1998)\citenamefont{Vescoli, Degiorgi,
  Henderson, Gr{\"u}ner, Starkey, and
  Montgomery}}]{vescoli_confinement_science}
\bibinfo{author}{\bibfnamefont{V.}~\bibnamefont{Vescoli}},
  \bibinfo{author}{\bibfnamefont{L.}~\bibnamefont{Degiorgi}},
  \bibinfo{author}{\bibfnamefont{W.}~\bibnamefont{Henderson}},
  \bibinfo{author}{\bibfnamefont{G.}~\bibnamefont{Gr{\"u}ner}},
  \bibinfo{author}{\bibfnamefont{K.~P.} \bibnamefont{Starkey}},
  \bibnamefont{and} \bibinfo{author}{\bibfnamefont{L.~K.}
  \bibnamefont{Montgomery}}, \bibinfo{journal}{Science}
  \textbf{\bibinfo{volume}{281}}, \bibinfo{pages}{1191} (\bibinfo{year}{1998}).

\bibitem[{\citenamefont{Schulz}(1996)}]{schulz_coupled_spinchains}
\bibinfo{author}{\bibfnamefont{H.~J.} \bibnamefont{Schulz}},
  \bibinfo{journal}{Phys. Rev. Lett.} \textbf{\bibinfo{volume}{77}},
  \bibinfo{pages}{2790} (\bibinfo{year}{1996}).

\bibitem[{\citenamefont{Donohue and
  Giamarchi}(2001)}]{donohue_deconfinement_bosons}
\bibinfo{author}{\bibfnamefont{P.}~\bibnamefont{Donohue}} \bibnamefont{and}
  \bibinfo{author}{\bibfnamefont{T.}~\bibnamefont{Giamarchi}},
  \bibinfo{journal}{Phys. Rev. B} \textbf{\bibinfo{volume}{63}},
  \bibinfo{pages}{180508(R)} (\bibinfo{year}{2001}).

\bibitem[{\citenamefont{Olshanii}(1998)}]{olshanii_potential_bec}
\bibinfo{author}{\bibfnamefont{M.}~\bibnamefont{Olshanii}},
  \bibinfo{journal}{Phys. Rev. Lett.} \textbf{\bibinfo{volume}{81}},
  \bibinfo{pages}{938} (\bibinfo{year}{1998}).

\bibitem[{\citenamefont{Cazalilla et~al.}(2003)\citenamefont{Cazalilla, Ho, and
  Giamarchi}}]{ho_dimcross_long}
\bibinfo{author}{\bibfnamefont{M.~A.} \bibnamefont{Cazalilla}},
  \bibinfo{author}{\bibfnamefont{A.~F.} \bibnamefont{Ho}}, \bibnamefont{and}
  \bibinfo{author}{\bibfnamefont{T.}~\bibnamefont{Giamarchi}}
  (\bibinfo{year}{2003}), \eprint{in preparation}.

\bibitem[{\citenamefont{Efetov and Larkin}(1975)}]{efetov_coupled_bosons}
\bibinfo{author}{\bibfnamefont{K.~B.} \bibnamefont{Efetov}} \bibnamefont{and}
  \bibinfo{author}{\bibfnamefont{A.~I.} \bibnamefont{Larkin}},
  \bibinfo{journal}{Sov. Phys. JETP} \textbf{\bibinfo{volume}{42}},
  \bibinfo{pages}{390} (\bibinfo{year}{1975}).

\bibitem[{\citenamefont{Zamolodchikov}(1995)}]{zamolodchikov_energy_sg}
\bibinfo{author}{\bibfnamefont{A.~B.} \bibnamefont{Zamolodchikov}},
  \bibinfo{journal}{Int. Review of Modern Physics A}
  \textbf{\bibinfo{volume}{10}}, \bibinfo{pages}{1125} (\bibinfo{year}{1995}),
  \eprint{and references therein}.

\bibitem[{\citenamefont{Thacker}(1981)}]{thacker_rmp}
\bibinfo{author}{\bibfnamefont{H.~B.} \bibnamefont{Thacker}},
  \bibinfo{journal}{Rev. Mod. Phys.} \textbf{\bibinfo{volume}{53}},
  \bibinfo{pages}{253} (\bibinfo{year}{1981}).

\bibitem[{\citenamefont{Donohue}(2001)}]{donohue_thesis}
\bibinfo{author}{\bibfnamefont{P.}~\bibnamefont{Donohue}}, Ph.D. thesis,
  \bibinfo{school}{Paris XI University} (\bibinfo{year}{2001}).

\bibitem[{\citenamefont{Pitaevskii and Stringari}(2003)}]{pitaevskii_becbook}
\bibinfo{author}{\bibfnamefont{L.}~\bibnamefont{Pitaevskii}} \bibnamefont{and}
  \bibinfo{author}{\bibfnamefont{S.}~\bibnamefont{Stringari}},
  \emph{\bibinfo{title}{Bose-Einstein Condensation}}
  (\bibinfo{publisher}{Clarendon Press}, \bibinfo{address}{Oxford},
  \bibinfo{year}{2003}).

\bibitem[{\citenamefont{Pedri and Santos}(2003)}]{pedri_tonks_harmonic}
\bibinfo{author}{\bibfnamefont{P.}~\bibnamefont{Pedri}} \bibnamefont{and}
  \bibinfo{author}{\bibfnamefont{L.}~\bibnamefont{Santos}},
  \bibinfo{journal}{Phys. Rev. Lett.} \textbf{\bibinfo{volume}{91}},
  \bibinfo{pages}{110401} (\bibinfo{year}{2003}).

\bibitem[{\citenamefont{{van Otterlo {\it et
  al.}}}(1995)}]{vanotterlo_2djjarrays}
\bibinfo{author}{\bibfnamefont{A.}~\bibnamefont{{van Otterlo {\it et al.}}}},
  \bibinfo{journal}{Phys. Rev. B} \textbf{\bibinfo{volume}{52}},
  \bibinfo{pages}{16176} (\bibinfo{year}{1995}).

\bibitem[{\citenamefont{{St{\"o}ferle {\it et
  al.}}}(2004)}]{Stoferle04}
\bibinfo{author}{\bibfnamefont{A.}~\bibnamefont{{{St{\"o}ferle\it et al.}}}},
  \bibinfo{journal}{Phys. Rev. Lett.} \textbf{\bibinfo{volume}{92}},
  \bibinfo{pages}{130403} (\bibinfo{year}{2004}).


\end{thebibliography}

\end{document}